\documentclass[apjl]{emulateapj}
\usepackage{graphicx}
\usepackage{color}
\usepackage{amsmath}

\shortauthors{Barclay, et al. 2013}
\shorttitle{A super-Earth-sized planet orbiting in or near the habitable zone around a Sun-like star}

\begin{document}

%
\def\ltsima{$\; \buildrel < \over \sim \;$}
\def\lsim{\lower.5ex\hbox{\ltsima}}
\def\gtsima{$\; \buildrel > \over \sim \;$}
\def\gsim{\lower.5ex\hbox{\gtsima}}
\def\K{\emph{Kepler}}

\newcommand{\numax}{\mbox{$\nu_{\rm max}$}}
\newcommand{\Dnu}{\mbox{$\Delta \nu$}}
\newcommand{\dnu}[1]{\mbox{$\delta \nu_{#1}$}}
\newcommand{\muHz}{\mbox{$\mu$Hz}}
\newcommand{\kep}{\mbox{\textit{Kepler}}}
\newcommand{\teff}{\mbox{T_{\rm eff}}}
\newcommand{\msun}{\mbox{$M_{\sun}$}}
\newcommand{\rsun}{\mbox{$R_{\sun}$}}
\newcommand{\lsun}{\mbox{$L_{\sun}$}}
\newcommand{\rearth}{\mbox{$R_{\oplus}$}}

\newcommand{\rp}{\mbox{$1.68^{+0.28}_{-0.28}$}}
\newcommand{\teq}{\mbox{$283\pm17$}}

\newcommand{\rpone}{\mbox{$2.67^{+0.45}_{-0.44}$}}

\newcommand{\name}{\mbox{Kepler-69 }}
\newcommand{\pone}{\mbox{Kepler-69b }}
\newcommand{\ptwo}{\mbox{Kepler-69c }}

%

\title{A super-Earth-sized planet orbiting in or near the habitable zone around Sun-like star}

\author{
Thomas Barclay$^{1,2}$,
Christopher J. Burke$^{1,3}$,
Steve B. Howell$^{1}$,
Jason F. Rowe$^{1,3}$,
Daniel Huber$^{1}$,
Howard Isaacson$^{4}$,
Jon M. Jenkins$^{1,3}$,
Rea Kolbl$^{4}$,
Geoffrey W. Marcy$^{4}$,
Elisa V. Quintana$^{1,3}$,
Martin Still$^{1,2}$,
Joseph D. Twicken$^{1,3}$,
Stephen T. Bryson$^{1}$,
William J. Borucki$^{1}$,
Douglas A. Caldwell$^{1,3}$,
David Ciardi$^{5}$,
Bruce D. Clarke$^{1,3}$,
Jessie L Christiansen$^{1,3}$,
Jeffrey L. Coughlin$^{1,3}$,
Debra A. Fischer$^{6}$,
Jie Li$^{1,3}$,
Michael R. Haas$^{1}$,
Roger Hunter$^{1}$,
Jack J. Lissauer$^{1}$,
Fergal Mullally$^{1,3}$,
Anima Sabale$^{7}$,
Shawn E. Seader$^{1,3}$,
Jeffrey C. Smith$^{1,3}$,
Peter Tenenbaum$^{1,3}$,
AKM Kamal Uddin$^{7}$,
and Susan E. Thompson$^{1,3}$
}

\newcommand{\ron}{}
\newcommand{\new}{}

\journalinfo{Accepted version, April 16, 2013}
\slugcomment{Accepted to ApJ}

\altaffiltext{1}{NASA Ames Research Center, M/S 244-30, Moffett Field, CA 94035, USA}
\altaffiltext{2}{Bay Area Environmental Research Institute, 596 1st Street West, Sonoma, CA 95476, USA}
\altaffiltext{3}{SETI Institute, 189 Bernardo Ave, Mountain View, CA 94043, USA}

\altaffiltext{4}{Department of Astronomy, University of California at Berkeley, Berkeley, California 94720, USA}
\altaffiltext{5}{NASA Exoplanet Science Institute, California Institute of Technology, 770 South Wilson Avenue, Pasadena, CA 91125, USA}
\altaffiltext{6}{Department of Astronomy, Yale University, New Haven, CT 06520, USA}
\altaffiltext{7}{Orbital Sciences Corporation/NASA Ames Research Center, Moffett Field, CA 94035, USA}

\begin{abstract}

We present the discovery of a super-earth-sized planet in or near the habitable zone of a sun-like star. The host is Kepler-69, a 13.7 mag G4V-type star. We detect two periodic sets of transit signals in the three-year flux time series of Kepler-69, obtained with the Kepler spacecraft. Using the very high precision Kepler photometry, and follow-up observations, our confidence that these signals represent planetary transits is $>$99.1\%. The inner planet, Kepler-69b, has a radius of $2.24^{+0.44}_{-0.29}$ \rearth and orbits the host star every 13.7 days. The outer planet, Kepler-69c, is a super-Earth-size object with a radius of $1.7^{+0.34}_{-0.23}$ \rearth and an orbital period of 242.5 days. Assuming an Earth-like Bond albedo, Kepler-69c has an equilibrium temperature of $299\pm19$ K, which places the planet close to the habitable zone around the host star. This is the smallest planet found by Kepler to be orbiting in or near habitable zone of a Sun-like star and represents an important step on the path to finding the first true Earth analog. 

\end{abstract}

\keywords{planetary systems; stars: fundamental parameters; stars: individual (Kepler-69, KIC 8692861, KOI-172); techniques: photometric; stars: statistics}

\section{Introduction}

NASA's Kepler mission was launched in 2009 with the primary aim of determining the abundance of Earth-size planets in our Galaxy \citep{borucki10}. The 0.95-m optical telescope is in an Earth-trailing heliocentric orbit, providing a stable thermal environment for the production of very high precision photometry. It comprises 42 CCDs with a combined field of view of 115 deg$^{2}$ \citep{koch10}. The brightness of over 160,000 stars is measured in 29.4-min integrations, and these data are searched for transiting planets, which appear as periodic dimming in brightness. 

While the first true Earth-analog, a planet the size of the Earth and in the habitable zone of a Sun-like star, has yet to be discovered, several important milestones towards achieving this goal have been met using data from the Kepler spacecraft. The first planet found by Kepler to be orbiting in the habitable zone of a Sun-like star was Kepler-22b \citep{borucki12}. With a radius of 2.4 $R_{\oplus}$, Kepler-22b likely does not have a rocky surface and is possibly a temperate ocean world \citep{rogers10}. Kepler has also been used to discover planets that are Earth-size \citep{batalha10,fressin12} and smaller \citep{muirhead12,barclay13} but these planets are likely to be too hot to host liquid water at their surfaces. The three catalogs of planet candidates published by the Kepler mission \citep{borucki11a,borucki11,batalha12} have detailed over 2300 candidates, and have each contained progressively smaller candidates in longer orbits. However, the observation timespan searched in each catalog (43 days, 137 days and 1.5 years, respectively) have thus far been too short to allow for three transits (necessary for period confirmation) of a potentially rocky planet in the habitable zone of a sun-like star.

Kepler-69 has been observed for nearly three years by the Kepler spacecraft (observational quarters Q1-Q12) in both long cadence mode (continuous 29.4-min integrations), and during Q3-Q6 and one month of Q7 in short cadence mode (continuous 58.8-sec integrations). Transit-like signals in the flux time series, occurring every 13.7-days with a depth of 600-ppm, have been previously reported \citep{borucki11,tenenbaum12,batalha12}. This signal was classified as a Kepler Object of Interest (KOI) with catalog number KOI-172.01. No other planet candidates were detected in the first 1.5 years of observations. Here we report the detection of a second planet candidate, KOI-172.02, with a period of 242.5 days, which was detected in a search of the first 2 years of observations. Since planets of this size and period cannot currently have their masses measured (by, for instance, radial velocity measurements) and their planetary nature thus confirmed, we rely on statistical analysis of the likelihood that KOI-172.01 and KOI-172.02 are planets for validation of their planetary status.

\section{Stellar properties}
\name appears in the Kepler Input Catalog \citep[KIC,][]{Brown11} as KIC 8692861 and has a magnitude of 13.7 in the Kepler bandpass. To derive stellar properties we obtained a high resolution spectrum of Kepler-69 using the HIRES spectrograph on the Keck~I telescope on 23 June 2011 using the setup of the California Planet Search group \citep{marcy08}.

The stellar properties of \name were determined by performing a $\chi^2$ fit of the spectrum with a library of 750 observed spectra of F--M type stars that have accurate parallaxes. The library spectra have effective temperatures spanning, $T_{\textrm{eff}}$ = 3500K--7500 K and surface gravities, $\log{g}$ = 2.0--5.0. The weighted mean of the ten library spectra with the lowest $\chi^2$ values were adopted as effective temperature, stellar surface gravity and metallicity of Kepler-69. The weights used were inversely proportional to the Pythagorean distance of stellar parameters from the median value, leading to an iteration that converges quickly. Choosing 10 library matches from which to determine the weighted average offers a statistical buffer against outliers. Best-fitting stellar properties are provided in Table~\ref{tab:stellar}.

We also ran an SME analysis \citep{valenti05} of the Keck spectrum of Kepler-69. The SME results were in excellent agreement our spectrum matching technique, albeit with lower uncertainties. However, following \citet{torres12} we inflated the uncertaities derived from the SME analysis to account for systematic biases. These inflated uncertainties were similar to those found from the spectral matching method described above so we chose to use spectral matching derived parameters in the remaining analysis.

We searched Kepler short cadence data for solar-like oscillations using the method of \citet{huber09} but did not detect any. Since oscillation 
amplitudes scale with stellar luminosity \citep{KB95}, evolved subgiant and giant stars show 
oscillations that are readily detectable with Kepler data of 13.7 mag stars. The 
non-detection of oscillations confirms that \name is a dwarf star with a lower limit of $\log{g}\gtrsim3.8$, consistent with the stellar properties derived from spectroscopy.

We matched the spectroscopically derived
temperature, surface gravity and metallicity 
to a fine grid of evolutionary models from the BASTI database \citep{basti} in order to estimate the star's mass and radius. 
We report the median and the 1-$\sigma$ region of each parameter derived from 
Monte-Carlo simulations using the spectroscopic values in Table~\ref{tab:stellar}.
Consistent parameters are found using Yonsei-Yale \citep{yy} and Dartmouth \citep{dartmouth} 
evolution models. \name is a main-sequence G4V star, 
with a somewhat lower mass and metallicity than the Sun.

\begin{table}
\begin{center}
\caption{Stellar properties of Kepler-69}
\vspace{0.1cm}
\begin{tabular}{l c c}        
\hline         
Parameter & Value\\  
\hline
$T_{\textrm{eff}}$ (K)&$5638\pm168$\\
$\log{g}$ (dex, cgs units)&$4.40\pm0.15$\\
Metalicity [Fe/H]&$-0.29\pm0.15$\\
Mass (\msun)&$0.810^{+0.090}_{-0.081}$\\
Radius (\rsun)&$0.93^{+0.18}_{-0.12}$\\
Luminosity (\lsun)&$0.80^{+0.37}_{-0.22}$\\
Density (g cm$^{-3}$)&$1.37^{+0.81}_{-0.55}$\\
Radial velocity zeropoint (km s$^{-1}$)&$-38.7\pm 0.1$\\
\hline
\end{tabular} 
\label{tab:stellar} 
\end{center}
\end{table}

\section{Transit analysis}
\subsection{Transit detection by the Kepler pipeline}
The Transiting Planet Search \citep[TPS,][]{jenkins10tps,tenenbaum12} and Data Validation \citep[DV,][]{wu10} modules of the Kepler Pipeline \citep{jenkins10} identified two transit-like signatures in a search of Q1--Q10 data. The first signature was due to KOI-172.01, a planet candidate previously reported by \citet{borucki11} and \citet{batalha12}. The second signature was new and had a period of $242.4579\pm0.0056$ days and a depth of $258\pm19$ ppm and has been designated KOI-172.02 (Burke et al. in prep.). The planet radius provided by the Q1-Q12 DV report for KOI-172.02 was 1.4 \rearth. A signal-to-noise ratio (SNR) of 73 was calculated for the transits of KOI-172.01 and 15 for KOI-172.02. The SNRs for both planet candidates are significantly above the formal threshold of 7.1 \citep{jenkins02}, giving us very high confidence that neither of these detections are due to random or correlated noise.

\subsection{Light curve preparation}
There are instrumental features near two of the KOI-172.02 transits. One was a sudden pixel sensitivity dropout due to a cosmic ray hit \citep{stumpe12}, the other is a coronal mass ejection \citep{DRNQ12}. Both aperture photometry and pre-search data conditioned \citep[PDC,][]{stumpe12,smith12} flux time series data stored at the Mikulski Archive for Space Telescopes (MAST) insufficiently corrected the out-of-transit flux levels around the instrumental signals. This led to a measured transit depth that was underestimated by 20\%. We instead chose to use a developmental version of the PDC error-corrected data that relies on error correction via wavelet-based band splitting \citep{stumpe12b}. This newly corrected data preserves stellar variability. We removed stellar variability using a numerically efficient discreet cosine transform \citep{garcia10} to implement a non-parametric penalized least squares smoothing of the flux time series. Data from individual quarters were normalized and then combined. Transits were treated as missing when smoothing the data. 

\subsection{Transit fitting}
We performed a fit of both planet candidates simultaneously using a \citet{mandel02} transit model with quadratic limb darkening. Limb darkening parameters were computed by interpolating the tables of \citet{claret11} and kept fixed.

The parameters included in our fit were: mean stellar density, photometric zeropoint,  and for each planet the transit epoch, orbital period, the planet to star radius ratio, impact parameter, and eccentricity vectors $e\sin{\omega}$ and $e\cos{\omega}$, where $e$ and $\omega$ are the eccentricity and argument of periastron of the orbit, respectively. We utilized the \texttt{emcee} implementation \citep{foreman12} of an affine-invariant Markov Chain Monte Carlo (MCMC) algorithm\citep{goodman10} to calculate posterior distributions for these parameters. The mean stellar density measured from asteroseismology was used as a Gaussian prior, all other parameters were given uniform priors except for the two eccentricity vectors. Using the $e\sin{\omega}$ and $e\cos{\omega}$ parameterization results in an implicit linear prior in $e$ \citep{ford06,burke08,eastman12}. We correct for this by enforcing a $1/e$ prior on eccentricity. Our likelihood function is therefore
\begin{equation}
\mathcal{L} = \exp{\left(\frac{\chi^{2}}{2} + \frac{\left(\rho_{a} - \rho_{m}\right)^{2}}{\sigma_{\rho}^{2}} - \ln{e} \right)},
\end{equation}
where $\rho_{a}$ is the mean stellar density derived from asteroseismology and $\rho_{a}$ is the model mean stellar density. $\chi^{2}$ is the usual sum of squared deviation from the model weighted by the variance.

The median and central 68\% of the posterior distribution (equivalent to the 1-$\sigma$ uncertainty) of each parameter are shown in Table~\ref{tab:transitparam}. Uncertainty in the stellar radius was included as a component of these calculations. This analysis yielded planet radii of $2.24^{+0.44}_{-0.29}$ and $1.71^{+0.34}_{-0.23}$ \rearth{} for KOI-172.01 and KOI-172.02, respectively.

\begin{figure}
\includegraphics[width=0.45\textwidth]{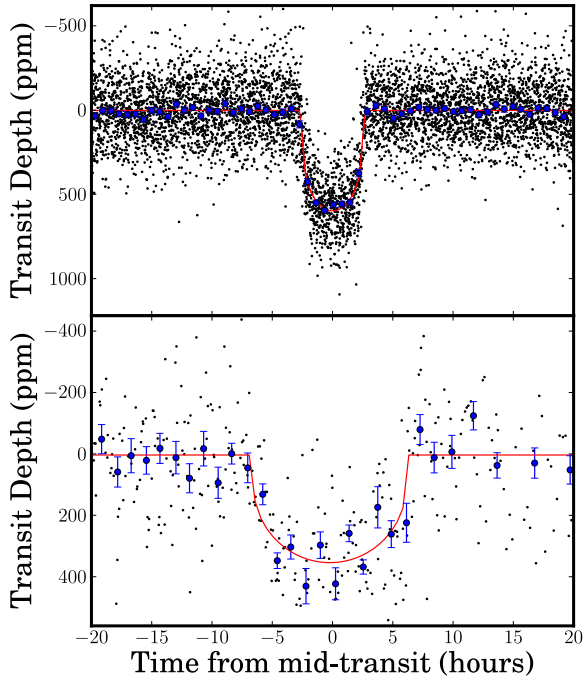}
\caption{The Kepler light curve of KOI-172. The upper panel shows the transits of KOI-172.01 folded on the orbital period and the lower planet shows the transits of KOI-172.02. The black points show the observed data and the blue points are binned data. In the upper panel 100 observed points are included in each bin and in the lower panel 12 point are included in each bin. The uncertainty on the binned data is the standard deviation divided by the square root of the number of point included in the bin. The best fitting transit models are shown overplotted in red.} 
\label{fig:transit}
\end{figure}

\begin{table*}
\begin{center}
\caption{Parameters from MCMC analysis}
\vspace{0.1cm}
\begin{tabular}{l c c}        
\hline         
Parameter & Kepler-69b (KOI-172.01)&Kepler-69c (KOI-172.02)\\  
\hline
Stellar density (g cc$^{-3}$)&\multicolumn{2}{c}{$1.05^{+0.48}_{-0.27}$}\\
$R_{p}/R_{\star}$&$0.02207^{+0.00023}_{-0.00018}$&$0.0168^{+0.00052}_{-0.00052}$\\
Mid-point of first transit (BJD - 2454833)\footnote{BJD-2454833 is the standard Kepler time system, known informally as BKJD.}&$137.8414^{+0.0016}_{-0.0016}$&$150.870^{+0.016}_{-0.014}$\\
Period (days)&$13.722341^{+0.000035}_{-0.000036}$&$242.4613^{+0.0059}_{-0.0064}$\\
Impact parameter&$0.15^{+0.17}_{-0.10}$&$0.12^{+0.21}_{-0.09}$\\
$e\cos{\omega}$&$0.02^{+0.19}_{-0.14}$&$-0.01^{+0.14}_{-0.16}$\\
$e\sin{\omega}$&$-0.07^{+0.09}_{-0.14}$&$-0.02^{+0.08}_{-0.15}$\\
Planet radius (\rearth)&$2.24^{+0.44}_{-0.29}$&$1.71^{+0.34}_{-0.23}$\\
$a/R_{\star}$&$21.8^{+2.9}_{-2.1}$&$148^{+20}_{-14}$\\
Semimajor-axis (au)&$0.094^{+0.023}_{-0.016}$&$0.64^{+0.15}_{-0.11}$\\
Inclination (deg)&$89.62^{+0.26}_{-0.45}$&$89.85^{+0.03}_{-0.08}$\\
Eccentricity& $0.16^{+0.17}_{-0.0010}$ & $0.14^{+0.18}_{-0.10}$\\
Equilibrium Temperature (K)\footnote{Equilibrium temperature assumes an albedo of 0.3 and the thermal recirculation parameter of 1.0.}& $779^{+50}_{-51}$ &$299^{+19}_{-20}$\\
Limb darkening parameters&\multicolumn{2}{c}{\{0.4012, 0.2627\}}\\
\hline
\end{tabular} 
\label{tab:transitparam} 
\end{center}
\end{table*}

\section{Tests to assess the validity of interpreting KOI-172.01 and KOI-172.02 as planets}
\subsection{Tests on the Kepler flux times series data}
The Kepler photometric data for KOI-172.01 and KOI-172.02 were examined for evidence suggestive of a false positive in the manner described in \citet{batalha12}. No such evidence was found and hence both candidates were afforded a place within the most recent Kepler planet candidate list\footnote{The Kepler planet catalog and DV reports are hosted at the NASA Exoplanet Archive, \url{http://exoplanetarchive.ipac.caltech.edu/}.} (Burke et al. in prep.). Specifically, odd and even numbered transits have consistent depths; there is no sign of a secondary transit; and there is no significant shift in the photo-centroid of the star in transit relative to out of transit \citep{jenkins10b}.

The Data Validation (DV) module of the Kepler pipeline creates three images; an average out-of-transit image taken from nearby but not during the transit events, an average in-transit image and a difference image which is the difference between the out-of-transit image and the in-transit image. A fit of the pixel response function (PRF) to the difference image gives the position of the transit source while a fit of the PRF to the out-of-transit image yields the position of the target star. The offset between the difference image position and the target star position can be used to identify false positive scenarios. Conversely, if no significant offset is detected, the uncertainty in the difference image centroid position can be used to calculate a radius of confusion outside of which a false positive source can be excluded. No significant offsets are measured between the difference image and target position for KOI-172.01. The 3-$\sigma$ radius of confusion was measured by DV at 0.16 arcsec. Only two transits were included in the PRF centroid offset metric in the DV report for KOI-172.02 owing to data artifacts and transits of KOI-172.01 occurring near the transits of KOI-172.02. We wished to use all five transits of KOI-172.02 in our centroid analysis so we created difference images in the same manner used by the Kepler pipeline \citep{bryson13} but included all observed transits of KOI-172.02. The significance of the offset from the remaining images was less than 3-$\sigma$ which is used as the warning threshold \citep{batalha12,bryson13}. The radius of confusion was estimated to be 1.5 arcsec. Given the above described tests, there is no reason to suspect KOI-172.01 or KOI-172.02 are false positives based on the Kepler data.

We measured a transit depth for KOI-172.02 of 350 ppm from the MCMC analysis. The faintest a star could be and still produce a transit of this depth can be calculated if we assume a total eclipse of that background star \citep{chaplin13}. Under this assumption the faintest a false positive star could be is 7.9 mag fainter than Kepler-69. This is the limit we use in all false positive calculations for KOI-172.02. For KOI-172.01 the transit depth is 597 ppm, which sets the maximum brightness difference between the target star and a false positive of 7.3 mag.

\subsection{Imaging of Kepler-69}
We obtained optical $V$-band images of Kepler-69 using the Nickel 1-m telescope at Lick Observatory and searched for nearby stars. No stars are seen 2--5 arcsec from Kepler-69 brighter than 19th magnitude; stars closer than 2 arcsec could not be ruled out.  Kepler-69 has a $V$-band magnitude of 14.0 \citep{everett12}, therefore we cannot exclude stars between V=19--21.9 for KOI-172.02 and V=19--21.3 for KOI-172.01 based on the Lick image and the transit depths.

\begin{figure*}
\includegraphics[width=0.95\textwidth]{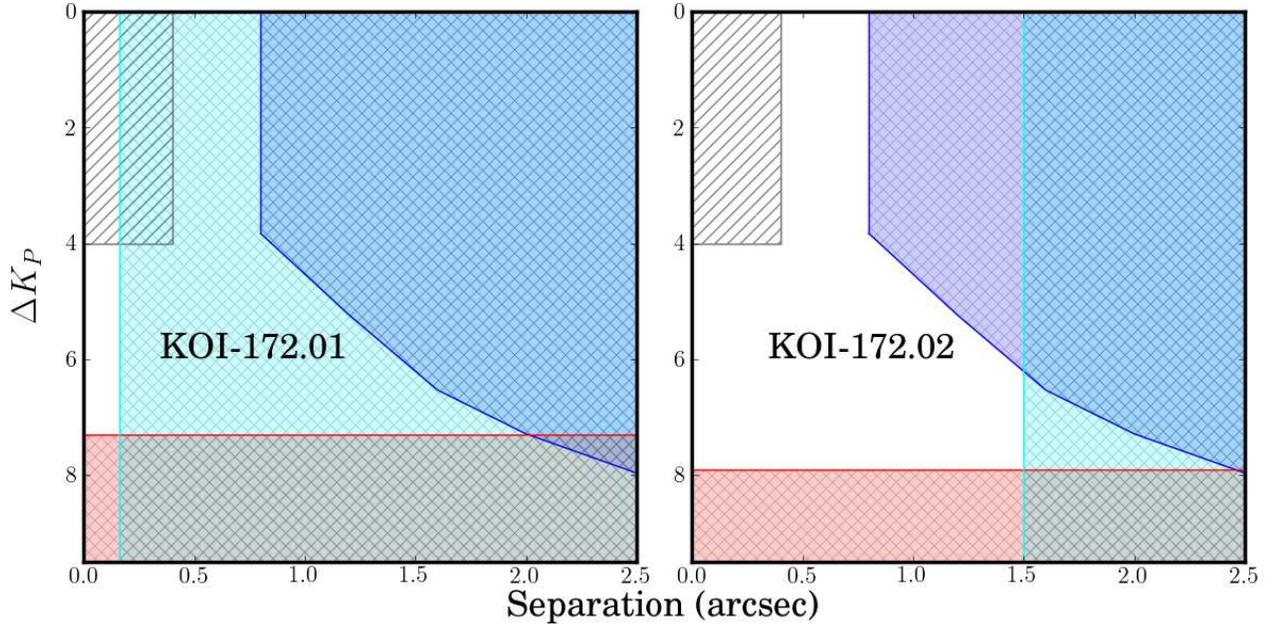}
\caption{The regions of parameter space where a false positive source can be excluded are shown colored. The exclusion region from the transit depth is shown in red, from the UKIRT imaging in blue and from the centroids in cyan. The white hashed region shows the constraints from the Keck spectral analysis. In this region contaminating stars are only allowed if they have a radial velocity within 10km s$^{-1}$. The region in white is not excluded and must therefore be accounted for in our false positive analysis.} 
\label{fig:exclusion}
\end{figure*}

There is a UK Infrared Telescope (UKIRT) $J$-band image of a field containing Kepler-69. The closest star is 3.2 arcsec west of Kepler-69 and is 9 mag fainter. This close-by star can be ruled out as the source of the transit signal from either planet candidate based on the PRF centroids. A magnitude dependent radius of confusion around Kepler-69 was calculated for use in false positive analysis. This confusion radius was converted to the Kepler magnitude system \citep{Brown11,howell12} and is shown in blue in Figure~\ref{fig:exclusion}.

\subsection{Looking for a second star in HIRES spectra}
The slit width used to collect the HIRES spectrum of Kepler-69 was 0.87 arcsec. This allowed us to search the HIRES spectrum for additional stars within 0.43 arcsec of the target. We calculated the best-fitting two-spectrum model using observed spectra with $T_{\textrm{eff}}$ ranging from 3500--6000 K. We searched over a range of secondary star radial velocities and flux ratios relative to the target star. No second star was seen in the spectra down to the confusion limit of $\Delta K_{P} =4.0$. We then injected a range of observed spectra into the spectrum of Kepler-69. Given we measure a radial velocity for Kepler-69 of $-38.7\pm0.1$ km s$^{-1}$, virtually all sufficiently bright background stars would have been detected. G and K dwarf physical companions to Kepler-69 that are further than approximately 20 au could evade detection owing to them having a radial velocity that we could not distinguish from Kepler-69. Companions with spectral type later than M0 are detected at any distance from the target (subject to falling into the spectrograph's slit) provided they are brighter than $\Delta K_{P} =4.0$.

\section{On the probability that KOI-172.01 and KOI-172.02 are false positives}
In this section we first consider the probability that KOI-172.01 and KOI-172.02 are in fact transits or eclipses of a background star. We find that the chance of this occurring for KOI-172.01 is 0.01\% and for KOI-172.02 it is 0.04\%. We are able to rule out the transit-like signals being due to a heirarchical triple stellar system and therefore validate KOI-172.01 and KOI-172.02 as confirmed planets.

We then go on to look at whether either of the two planets orbit a physical companion star. While for KOI-172.01 the probability of this is 0.01\%, for KOI-172.02 it is $<$1\% and hence cannot be definitively ruled out, but is unlikely.

\subsection{False positive scenarios considered}
Given that the KOI-172.01 and KOI-172.02 detections are unlikely to be due to noise \citep{jenkins02}, we can safely assume that they either represent planets or astrophysical false positives. Scenarios that could cause a transit-like signal that are not caused by a planet orbiting \name are:
\begin{enumerate}
\item Stellar binaries with grazing eclipses (the impact parameter $>$1).
\item Background eclipsing binaries\footnote{The phrase background stars is used to refer to stars that appear fainter than \name and are not physically associated. They may be either in the foreground or background with respect to Kepler-69.}.
\item Background star-planet systems.
\item An eclipsing binary physically associated with \name in a heirarchical stellar triple system.
\item A planet orbiting a stellar physical companion of Kepler-69.
\end{enumerate}

We are immediately able to dismiss scenarios 1 and 4 because the best-fitting models representing these scenarios fit the observed data worse than the best-fitting transiting planet model at a level $>$3$\sigma$. This is because a stellar eclipse across Kepler-69 or a physical companion would be more V-shaped than we observe. Additionally, these scenarios are predicted to be very uncommon for sub-Jupiter-sized planet candidates \citep{fressin13}. The remaining false positive discussion is restricted to scenarios 2, 3 and 5.

\subsection{Background false positive scenarios}
To assess the probability that the transit-like signals we detect are not associated with the Kepler-69 system but are on a background star, we simulate the stellar background population that could be responsible. We simulated the number of stars within 1 deg of Kepler-69 using the Besan\c{c}on Galaxy model with kinematics enabled \citep{robin03}. The Galactic model predicts there are 432,118 stars, of which 44,185 fall in the brightness range $14.0<V<22.0$ and are not excluded by the secondary star test performed on the Keck spectrum. We integrated the area shown in white in Figure~\ref{fig:exclusion} to calculate the undetected stellar population in the background of KOI-172.02. The estimated number of stars in the background is 0.018. If we now consider that only 2.6\% of stars observed by Kepler host either a transiting planet candidate or are an eclipsing binary, excluding contact binaries \citep[][Burke et al. in prep.]{batalha12,slawson11}, the predicted number of background stars which could cause a false positive is 0.00046.

We used a Bayesian approach similar to that applied by \citet{fressin12} and \citet{barclay13} to the validation of the planets orbiting Kepler-20 and Kepler-37 to quantify the false positive probability. This technique compares the \textit{a priori} chance of finding a planet the size of KOI-172.02 (the planet prior) to that of finding a background false positive. The ratio of the planet prior to the false positive probability is used to calculate our confidence in the planet interpretation. The population of super-earth size planets orbiting with periods longer than 85 days is not well constrained. We assume that the super-earth occurrence rate is the same at long orbital periods as it is at shorter orbital periods when period is measured in logarithmic units. There is evidence that the occurrence rate of super-Earth-size planets is even high at longer orbital periods than at short periods. For example \citet{cassan12} find $62^{+35}_{-37}$\% of stars host a super-Earth-mass planet. Therefore, using the statistics of \citet{fressin13} who derived a super-Earth occurrence rate of $23\pm2.4$\% may well lead to an underestimate of our planet prior.

There are 666 super-earth-size (1.25--2.0 \rearth) planet candidates tabulated in \citet{batalha12}. However, \citet{fressin13} predict that 8.8\% of these are false positives which provides an estimated number of super-earths found by Kepler of 607. The catalog of \citet{batalha12} was based on continuous (Q1--Q6) observations of 138,253 stars. Therefore, the occurrence rate of transiting super-Earth-size planets is 0.0044 per star. We can compare this to the background/foreground false positive population which is 0.00046, as explained above. If KOI-172.02 were the only planet candidate associated with Kepler-69, the likelihood of it being a true planet would be $0.0044 / (0.0044 + 0.00046) =0.91\,$. However, KOI-172.02 is in a two planet candidate system. As shown by \citet{lissauer12}, having multiple transiting planet candidates associated with a star increases the probability that these candidates are real planets by a factor of $\sim$15 compared to single planet candidate systems. This \emph{multiplicity boost} increases our confidence that KOI-172.02 is a \textit{bona fide} planet physically associated with the Kepler-69 system to 99.3\%. 

We note that Kepler-69 has a lower metallicity than the Sun and for large planets a low metal content has been shown to hinder planet occurrence \citep[e.g.][]{mayor11}. However, for there appears to be no link between planet occurrence and stellar metalicity \citep{buchhave12} for transiting super-Earth-size planets. Therefore we do not factor the stellar metalicity into our our false positive calculations.

The centroid confusion radius for KOI-172.01 is much smaller than for KOI-172.02 (as shown in Figure~\ref{fig:exclusion}). Using the approach outlined above, our confidence is 99.9\% that KOI-172.01 is a planetary mass body physically associated with the Kepler-69 system. We can therefore conclude that the signals we observe are due to \emph{bona fide} planets. The inner and outer planets are named Kepler-69b and Kepler-69c, respectively.

\subsection{The possibility that the planets in the Kepler-69 system orbit a physical companion.}
While the above analysis leads us to conclude that Kepler-69b and Kepler-69c are both sub-stellar bodies physically associated with the target system, this does not exclude a scenario where one or both of the planets orbit a fainter stellar companion to Kepler-69.

We performed a series of MCMC transit analyses with no prior on the stellar density with the aim of constraining the range of stellar types the planets could orbit. First we assumed that both planets orbit the same star. If this is the case then this star would have a stellar density of $1.04^{+0.53}_{-0.31}$ g cm$^{-3}$ with a 3-$\sigma$ upper limit of 3.6 g cm$^{-3}$. This rules out both planets orbiting a star with a radius less than 0.64 $R_{\sun}$, where the radius is calculated using Dartmouth isochrones \citep{dartmouth} and assumes Kepler-69 and stellar companion co-evolved. We tried including a fixed dilution in the transit depth of 98\% to simulate both planets being around a star 1/50th the brightness of the primary star in the system. In this case we found a 3-$\sigma$ upper limit on the stellar density of 2.8 g cm$^{-3}$. However, a companion hosting a transit that who's light is diluted by 98\% should have a density of 9.7 g cm$^{-3}$ based on isochrones, therefore we can rule out a scenario where both planets orbits a much fainter companion. If both planets orbit a companion, it must be a G or early K-type star.

We now consider the case that both planets orbit different stars. Fitting only the transits of Kepler-69b gave a stellar density upper limit of 6.2 g cm$^{-3}$ which would place Kepler-69b around a star with a radius of at least 0.48 $R_{\sun}$. This rules out Kepler-69c orbiting a star with a spectral type later than K9V. Fitting just for Kepler-69c implies a star with a mean stellar density of $1.3^{+1.1}_{-1.0}$ g cm$^{-3}$ and the 3-$\sigma$ upper limit on density is 8.7 g cm$^{-3}$. Including a dilution in the transit depth of 98\% loosened our $3-\sigma$ upper limit on the stellar density to 15.5 g cm$^{-3}$. A star with a density of 15.5 g cm$^{-3}$ that co-evolved with Kepler-69 would have a radius of 0.31 $R_{\sun}$ -- an M2-type star. Therefore, in the case of a physical companion, Kepler-69c cannot orbit a star cooler than an M2-type.

To quantify the probability that either of the planets orbit a companion star, we constructed three Monte Carlo simulations to estimate the properties of a stellar companion with different assumptions: both planets orbit a companion to Kepler-69; Kepler-69b orbits Kepler-69 and Kepler-69c orbits a companion; and Kepler-69c orbits Kepler-69 and Kepler-69b orbits a companion. 


In the simulations we assume the orbital period is described by a log-normal distribution with mean $\log{P} = 5.03$ and standard deviation $\sigma_{\log{P}} = 2.28$ where period is in days \citep{raghavan10} and eccentricity is uniform from zero to one. The primary to secondary mass ratio $q$ is assumed to be uniformly distributed \citep[this approximates the distribution found by][]{halbwachs03}. The argument of periastron is taken to be uniformly distributed. We ran the simulation 10$^{7}$ times. At each loop in the simulation we calculated the radius, density and brightness of the companion based on Dartmouth isochrones \citep{dartmouth}. We calculated the binary separation from the orbital period and mass ratio. The radial velocity at the point of observation of the secondary was determined by calculating a radial velocity curve and selecting the time of observation at random.

We then calculated the proportion of secondary scenarios from our simulation that are unphysical or excluded based on observations. Unphysical scenarios are those where the binary separation is too small to allow a stable planetary orbit of 242.4 days \citep{rabl88,holman99}. We used the diluted MCMC transit analyses' 3-$\sigma$ upper limit on stellar density to exclude companions that are too dense or faint. Using the constraint from the Keck spectrum, if a companion had a brightness within 4 mag of Kepler-69 and a radial velocity difference between it and the primary of $>$10 km s$^{-1}$ (the limit from our analysis of the Keck spectrum) it was excluded. Companions that would be seen by the J-band image or would induce a detectable centroid shift were also excluded. We find that a companion star is able to host Kepler-69b less than 0.01\% of the time. Therefore, we are confident that Kepler-69b orbits the target star. However, in 1.9\% of the simulations Kepler-69 has a stellar companion capable of hosting Kepler-69c. If we assume that in scenarios with a viable planet hosting companion, half the time the planet orbits the primary and half the secondary, then our confidence that Kepler-69c orbits the primary is 99.1\%. However, no \emph{multiplicity boost} is included in the above calculation. It is unclear whether a companion is more likely to host transiting planets if the primary hosts transiting planets compared to a field star. However, if we were to apply this boost, the probability that Kepler-69c orbits Kepler-69 is 99.93\%. The true probability that Kepler-69c orbits a companion star lies somewhere between the 0.06\% and 0.9\%. 

While we cannot conclusively rule out the possibility that Kepler-69c orbits a smaller star and therefore is larger and cooler, the chances are low. With more follow-up observations we may be able to further restrict the allowed parameter space and increase our confidence that Kepler-69c orbits the primary star. Specifically, continued non-detections in future radial velocity observations will increase the size of the region around Kepler-69 where we can rule out any companion star, while a deep adaptive optics observations can help rule out companions further from the star than around 0.5 arcsec. 

In the remaining analysis we will assume that Kepler-69c orbits the target star.

\section{On the composition and temperature of Kepler-69\lowercase{c}}
It is not clear where the boundary between the radius of a rocky (or terrestrial) planet and that of an giant volatile rich planet lies. There is very likely not an absolute boundary but a range of cutoffs which depend on the initial composition of the proto-planetary disk \citep{valencia07}. In any case, the 1-$\sigma$ uncertainty we determine for the radius of Kepler-69c makes predicting its composition difficult. The 1-$\sigma$ lower bound of 1.48 \rearth{} may well represent a rocky planet but the upper bound of 2.04 probably represents a fairly volatile rich planet. If Kepler-69c were volatile rich is may be a water world and quite unlike any planet in our Solar System.

\begin{figure}
\includegraphics[width=0.45\textwidth]{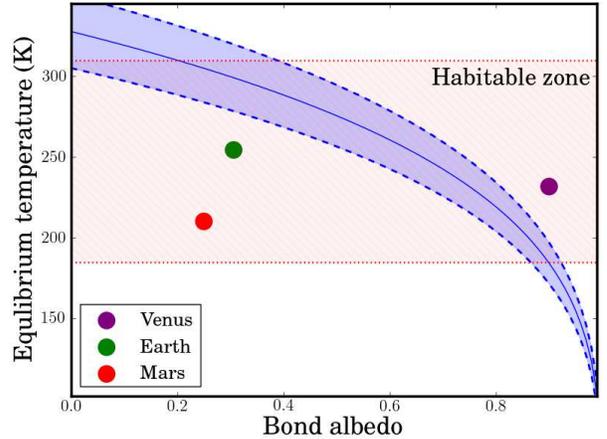}
\caption{The equilibrium temperature of Kepler-69c as a function of Bond albedo. The blue shaded region shows the 1-$\sigma$ range assuming $f=1.0$ where uncertainties were calculated from the MCMC chains. We expect $f$ to be close to unity if Kepler-69c has an atmosphere similar to that of Earth. Venus, Earth and Mars are plotted for comparison. The habitable zone as suggested by \citet{kasting11b} is shaded in red and bounded by the dashed lines. Kepler-69c falls near or within the habitable zone with reasonable assumptions of its albedo.} 
\label{fig:teq}
\end{figure}

To estimate the equilibrium temperature of Kepler-69c we must make some assumptions about the albedo and the thermal energy redistribution of the planet. We use the same equation as used in \citet{batalha12} to calculate equilibrium temperature,
\begin{equation}
T_{\textrm{eq}} = T_{\textrm{eff}}(R_{\star}/2a)^{1/2} [f(1-A_{B})]^{1/4},
\end{equation}
$T_{\textrm{eff}}$ is the effective temperature of the star, $R_{\star}$ is the stellar radius, $a$ is the planet's semi-major axis, $f$ is the thermal redistribution parameter (where $f=1$ refers to full thermal recircularization), and $A_{B}$ is the Bond albedo. In the \citet{batalha12} Kepler planet candidate catalog they assume $f=1$ and $A_{B}=0.3$ (the same albedo as Earth). Using these values results in a temperature of $299^{+19}_{-19}$ K, significantly lower than the boiling point of water. If we assume a lower Bond albedo of 0.1 (similar to that of Mercury) the temperature of the planet increases to $318^{+20}_{-21}$ K, still in a regime where water could exist in a liquid form (neglecting the effects of an atmosphere). In Figure~\ref{fig:teq} we show the range of equilibrium temperatures that Kepler-69c would have as a function of $A_{B}$. With a reasonably thick atmosphere and a rotation rate similar to Earth, Kepler-69c would have a thermal redistribution parameter of $f\sim 1$ (for reference Earth has a value of $f=1.1$ and for Venus $f\sim1$).

An atmosphere can dramatically increase the surface temperature of a planet (c.f. Venus). \citet{kasting11b} suggests the upper boundary on the temperature of a habitable planet lies around 309 K -- Kepler-69c likely falls below this limit if it has a similar or higher albedo than Earth. However, \citet{selsis07} puts the limit at a more conservative 270 K, Kepler-69c would require an albedo greater than 0.5 to satisfy this requirement.  On Figure~\ref{fig:teq} we have plotted the habitable range suggested by \citet{kasting11b}. We have also plotted points corresponding to Venus, Earth and Mars. Kepler-69c likely falls within the habitable zone for moderate albedos if we use the limit of \citet{kasting11b} although it would require a fairly high albedo to be habitable according to the limit of \citet{selsis07}. 

The above analysis assumes that Kepler-69c has a circular orbit. From our MCMC transit analysis we find Kepler-69c may be highly eccentric -- the 3-$\sigma$ upper limit is $e=0.79$. The apastron distance is $0.73^{+0.22}_{-0.13}$ au at which distance Kepler-69c would have an equilibrium temperature of $282^{+19}_{-31}$ K. However, Kepler-69c would have an equilibrium temperature of $322^{+47}_{-21}$ K at periastron. If the planet is indeed on an eccentric orbit it may experience high seasonal temperature variations. However, a Kepler-69c is most likely on a reasonably circular orbit and therefore will probably not go through dramatic freeze-thaw cycles.

\section{Conclusions}
In this paper we report the discovery of two transit-like signals in the light curve of Kepler-69 which we show are \textit{bona fide} planets orbiting a Sun-like star. While we are confident at the 99.9\% level that Kepler-69b orbits the target star, there is a slight possibility ($<0.9\%$) that Kepler-69c orbits a physical companion star to Kepler-69. 

With a 1-$\sigma$ uncertainty on the stellar radius and hence the radius of Kepler-69c of 17\% we cannot conclusively say whether the planet is rocky or has a high volatile content \citep{valencia07}. 

The temperature of Kepler-69c falls close to its host star's habitable zone but a degree of inference must be made as to the planet's albedo in order to determine an equilibrium temperature. With reasonable assumptions it is possible that liquid water could exist in significant quantities on the surface of this planet.

Kepler-69c represents the first discovery of a super-Earth in the habitable zone of a Sun-like star. The work represents a progressive step on the road to detecting the first truly Earth-like planet orbiting a star like our Sun.

\acknowledgements
This paper includes data collected by the \K{} mission. Funding for the \K{} mission is provided by the NASA Science Mission Directorate. Some \K{} data presented in this paper were obtained from the Mikulski Archive for Space Telescopes (MAST) at the Space Telescope Science Institute (STScI). STScI is operated by the Association of Universities for Research in Astronomy, Inc., under NASA contract NAS5-26555. Support for MAST for non-HST data is provided by the NASA Office of Space Science via grant NNX09AF08G and by other grants and contracts. This research has made use of the NASA Exoplanet Archive, which is operated by the California Institute of Technology, under contract with the National Aeronautics and Space Administration under the Exoplanet Exploration Program. We used data from the UKIRT Wide Field Camera \citep[WFCAM,][]{casali07} and a photometric system described in \citet{hewett06}. The pipeline processing and science archive are described in \citet{hambly08}. DH is supported by appointment to the NASA Postdoctoral Program at Ames Research Center, administered by Oak Ridge Associated Universities through a contract with NASA.


\end{document}